\documentclass[floatfix,rmp,twocolumn,twoside]{revtex4}
\usepackage{graphicx}

\bibpunct{[}{]}{,}{n}{}{,}

\usepackage{graphicx,times,amsmath, fontenc}
\usepackage{multirow}
\usepackage{afterpage}

\usepackage[english]{babel}

\makeatletter
\g@addto@macro\@verbatim\small
\makeatother

\begin{document}

\title{Community-Quality-Based Player Ranking in Collaborative Games with no Explicit Objectives}
\author{Luis~Quesada and Pablo~J.~Villacorta\\
  Department of Computer Science and Artificial Intelligence, CITIC, University of Granada, \\
  Granada 18071, Spain \\
  \textit{lquesada@decsai.ugr.es, pjvi@decsai.ugr.es}
}

\begin{abstract}
Player ranking can be used to determine the quality of the contributions of a player to a collaborative community.
However, collaborative games with no explicit objectives do not support player ranking, as there is no metric to measure the quality of player contributions.
An implicit objective of such communities is not being disruptive towards other players.
In this paper, we propose a parameterizable approach for real-time player ranking in collaborative games with no explicit objectives.
Our method computes a ranking by applying a simple heuristic community quality function.
We also demonstrate the capabilities of our approach by applying several parameterizations of it to a case study and comparing the obtained results.
\end{abstract}

\maketitle
\section{Introduction}

Player ranking in collaborative games allows determining whether a player is contributing to the community or harming it. Player rankings are used to give rewards fair players with perks and to punish disruptive players. These perks or penalties alter the reach of the player's actions so their influence on the community is proportional to how good the player's contributions are.

When the game has clear shared objectives, it is easy to determine how much a player is contributing to the community, just by measuring how much the player's actions contribute to the objective completion. This has been widely studied \cite{Shapley1953,Faigle1992,Huang2008,Brandt2009}.

However, when the game has no clear objectives, no metric exists to measure player contribution quality. Indeed, each player may have a different personal motivation to achieve different self-imposed goals \cite{Hainey2011}, and player actions can be considered fair or disruptive towards the community depending on whether they respect or damage other player contributions.
In these cases, there is a very abstract and subjective shared implicit objective that could be described as building a fair and not disruptive player community.
It should be noted that fair players benefit from their behavior, as it is more likely that other players act fair towards them.
Furthermore, a community of disruptive players seems to repel fair players and the community quality has an intuitive tendency to gradually drop off. Contrarily, a community of fair players lures new fair players, which lead, in turn, to an increase of the community quality.

As a player ranking can be used to fairly assign perks and penalties, and such alterations raise the quality of a community, a player ranking approach based on community quality closes a feedback loop and allows collaborative communities to self-control and gradually increase quality.

In this paper, we propose and compare a parameterizable approach for real-time player ranking in collaborative games with no explicit shared objectives that produces player rankings from the application of a simple, domain-specific heuristic function to measure the quality of the community.

Section \ref{sec:back} introduces real-world scenarios that can benefit from player ranking.
Section \ref{sec:rtpr} describes our approach to real-time player ranking in collaborative games with no explicit objectives.
Section \ref{sec:comparison} presents, evaluates, and compares results obtained from applying different parameterizations of our approach to a case study.
Section \ref{sec:concfw} exposes our conclusions and the future work that derives from our research.

\section{Background} \label{sec:back}

We now proceed to introduce several cases of collaborative games with no explicit objectives. We also explain how a player ranking allows the implementation of mechanisms that improve these communities.

Subsection \ref{sec:bmine} introduces a real case of collaborative game: Minecraft.
Subsections \ref{sec:bdforum} and \ref{sec:brsite} describe a discussion forum and a recommendation site, respectively, and model them as collaborative games.
Finally, Subsection \ref{sec:bccg} presents the collaborative clustering game, an abstraction of collaborative games with no explicit objectives that we will use to demonstrate the validity of our approach throughout this work.

\subsection{Minecraft} \label{sec:bmine}

Minecraft is a 3D open world video game \cite{Minecraft} in which players can collect resources pictured as textured cubes and arrange them to build contraptions.

Although Minecraft has no explicit objective, the fact that it supports multiplayer mode makes it a collaborative game: players may help each other gathering resources and building complex constructions. However, disruptive players may join an ongoing game play and deliberately destroy other player artifacts. Administrators can ban players but, in order to effectively do so, they should have to keep an eye on every single player in the community.

Player rankings in Minecraft would highlight disruptive players and help distinguish them from fair players that have accidentally caused slight damages to other player contraptions. It should be noted that such a ranking would allow the administrators to focus their attention on the potentially disruptive players.

\subsection{A Discussion Forum} \label{sec:bdforum}

A discussion forum is a site where people can hold conversations in the form of posted messages. Users taking part in them ask and answer questions to other users.

Forums have no explicit objectives. However, they have the implicit objective of putting together a reliable community that provides reliable answers to user inquiries.

Perks in forums involve increasing the visibility of fair player messages, while penalizations often involve banning disruptive users to deter or prevent them from bothering others \cite{Zhang2012}.

Player rankings in discussion forums would uncover disruptive behaviors. This information can be exploited by moderators, which would only need to focus their efforts in watching potentially disruptive users.

\subsection{A Recommendation Site} \label{sec:brsite}

A recommendation site seeks to predict user preferences by considering the user's item ratings and the complex social networks users themselves conform.

Recommendation sites, as discussion forums, have no explicit objectives. However, they have the implicit objective of training the system to provide good recommendations to users.

Perks and penalizations in recommendation sites are often reflected as user reputation that determine how much their ratings are taken into account \cite{Massa2004,ODonovan2005}.

Player rankings in recommendation sites would point out possibly troublesome players who only seek to destabilize the rating system, for example by incorrectly rating items. Such a player ranking can be used to alter player reputations and, in turn, the value of such unfair player ratings.

\subsection{The Collaborative Clustering Game} \label{sec:bccg}

As part of our research, we have developed the collaborative clustering game abstraction as a case study.

\begin{figure}[tb]
\centering
\includegraphics[scale=1]{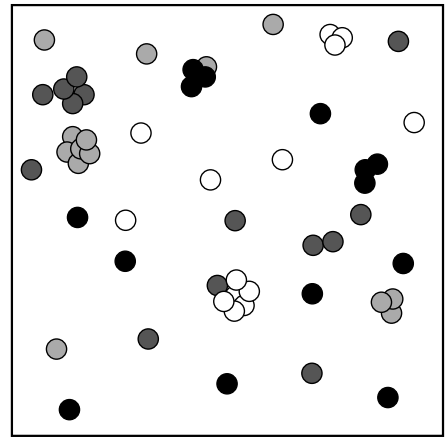}
\caption{A collaborative clustering game board}
\label{fig:colclus}
\end{figure}

The collaborative clustering game is a multi-player real-time collaborative video game. In it, all the players share a board with the same number of white, light gray, dark gray and black dots. Players can drag and drop any dot and the movement is mirrored in real time on every player screen. A dot that is being dragged cannot be moved by another player until dropped. An example of collaborative clustering game board is shown in Figure \ref{fig:colclus}.

We assume that players do not get points from their actions in this game (i.e. there is no explicit objective), but that they somehow benefit from having all the dots of the same color clustered together and apart from other color dots. For example, assume that users have the task to cluster any new dots that appear in the board, and the fact clusters are clearly distinguished benefits players in fulfilling their task.

It should be noted that this is a pseudo-game, an abstraction of game models similar to those for a discussion forum or a recommendation site: users interact with the community contents in real time, their actions modify the community quality, and a high quality community benefits the user. Also, there could be fair or disruptive players.

Measuring the quality of a collaborative clustering game is trivial and, therefore, we will use it throughout this paper to objectively evaluate the player ranking approach we present in the next section.

\section{Player Ranking Based on Community Quality} \label{sec:rtpr}

We proceed to explain our parameterizable approach for player ranking in collaborative games with no explicit objectives using a heuristic community quality function.

Subsection \ref{sec:pralgo} describes our expression for player contribution quality rating.
Subsection \ref{sec:prasym} exposes an asymptotic analysis of our algorithm.
Subsection \ref{sec:prtrain} explains our system training and evaluation capabilities.
Finally, Subsections \ref{sec:prmine} to \ref{sec:prccg} present examples of applications of our system to the games introduced in Section \ref{sec:back}.

\subsection{Contribution Quality Rating} \label{sec:pralgo}

Although obviously there does not exist an exact metric for the quality of a community, there can be defined heuristic functions that approximate the community quality value. Several examples of such heuristics will be studied further in this Section.

The contribution quality rating or CQR for a player can be estimated from the effect the player's actions have on the community quality.
Whenever a player's action causes the community quality to increase (resp. decrease), the player's CQR has to increase (resp. decreases).

While the player CQRs are not relevant on their own, a player ranking based on CQRs is a good measure of the quality of the contributions of each player relative to others in the community.
Automatic preventive actions such as penalizations or alarms can be triggered when a player's CQR meets criteria such as exceeding a threshold that is either absolute or relative to the distribution of the community CQRs.

Communities consists of a set of community quality domains. Each community quality domain can be assigned a different heuristic community quality function and can represent different partitions of the community. For example, it is possible to establish a correspondence between different demographic groups in a recommendation site and community quality domains, and it is possible to define measures of different phenomena such as use of capital letters or swear words in a discussion forum as community quality domains. It should be noted that these domains may even overlap, that is, a single player action may affect several of them.

Each community quality domain $C_i$ in a community $C$ is to be assigned a heuristic function $Q(\cdotp)$ that approximates its quality value.
The change caused to the quality of a community $C$ by a player when he performs an action $a$ is denominated delta value of the action and can be formalized as a function $\Delta(C,a)$. This function is the sum of the changes in the quality of all the community quality domains as caused by the player action:

\begin{equation} \label{eq:cqr1}
\Delta(C,a) = \sum_{j = 1}^{|C|} Q(a(C)_j) - Q(C_j)
\end{equation}

In the equation above, the effect of a player action $a$ on a community has been formalized as a function $a(\cdotp)$, that applies the action to a community and returns, as a result of the action, the resulting community.

As a first approach, given $\Delta$, the list of the $n$ delta values $\Delta = \{ \Delta(C, a_1), \Delta(a_1(C),a_2),$ $..., \Delta(a_{n-1}(...a_1(C)))), a_n) \}$ corresponding to a sequence of $n$ player actions, the CQR of a player is defined as:

\begin{equation} \label{eq:cqr1}
CQR(\Delta) = \sum_{i=1}^n \Delta_i
\end{equation}

This approach clearly has a major flaw: it is not readily responsive to changes in players' behavior. For example, a fair player may turn disruptive due to an argument with other players. It would take a long time until the recent disruptive action deltas surpass the past fair actions and, therefore, when using this expression it would take a long time to tag these players as disruptive. Intuitively, it is necessary to add some kind of temporal constraint to the expression. This way, a player that starts damaging the community can be quickly tagged as disruptive.

In order to solve this flaw, we propose a parameterization of our previous CQR expression, as follows:

\begin{equation} \label{eq:cqr2}
CQR_{T}(\Delta) = \sum_{i=n-T+1}^n \Delta_i
\end{equation}

When using this approach, only the latest $T$ deltas are considered in the calculation of a player's CQR. Now, when a player changes its behavior, it will only take up to $T/2$ actions on average to notice that the behavior has changed. The proposal in (\ref{eq:cqr1}) is, indeed, a particular case of this approach when $T = \infty$.

Nevertheless, this new approach has a new drawback. Series of low value positive actions may force the expression to ignore high value actions. 

In order to solve this drawback, define $f_x$, a filter that removes the deltas whose absolute value is less than $x$ in a list, and we propose a new parameterization of our previous CQR expression, as follows:

\begin{equation} \label{eq:cqr3}
CQR_{T,x}(\Delta) = \sum_{i=n-T+1}^n f_x(\Delta_i)
\end{equation}

When using this approach, only the latest $T$ deltas whose absolute value is greater than or equal to $x$ are considered in the calculation of a player's CQR. Insignificant actions will now be ignored and they will not mask relevant action deltas. The proposal in (\ref{eq:cqr2}) is, in turn, an instance of this more generic approach, with $x = 0$.

It should be noted that this filter brings back some responsiveness issues to our approach, as an average of $T/2$ actions are needed to detect a change in behavior.

In order to solve this problem, we define $g_k$, another filter that, whenever the last $k$ elements of the already filtered delta list are all positive (resp. negative), all the negative (resp. positive) values in the list are made into zeros. We then propose a final parameterization of our previous CQR expression, as follows:

\begin{equation} \label{eq:cqr4}
CQR_{T,x,k}(\Delta) = \sum_{i=n-T+1}^n g_k(f_x(\Delta_i))
\end{equation}

This new filter emulates the effect of the differential correction in a PID controller and, indeed, raises the responsiveness of the expression so that, apart from the $T/2$ average actions needed to detect a change in behavior, a maximum of $k$ actions is enough to detect it. The proposal in Equation \ref{eq:cqr3} is a particular case of this more generic approach, when $k = T$.

\subsection{Asymptotic Analysis} \label{sec:prasym}

The memory requirements of an efficient implementation of the CQR expression are:

\begin{equation}
mem(T,k,P) = \left\{ \begin{array}{ll}
               P, & \mbox{if $T=\infty$, $k=T$} \\
               (k+2)\cdot P, & \mbox{if $T=\infty$, $k<T$} \\
               T\cdot P, & \mbox{if $T<\infty$, $k=T$} \\
               (T+2)\cdot P, & \mbox{if $T<\infty$, $k<T$}
              \end{array} \right.
\end{equation}

\noindent being $P$ the number of players.

As $k$ is always less than or equal to $T$, in the worst case, $mem(T,k,P) = (T+2)\cdot P$. $T$ can be considered a constant, therefore $mem(T,k,P)$ is $O(P)$.

The processing time requirements of an efficient implementation of the CQR expression are:

\begin{equation}
time(T,k,A) = \left\{ \begin{array}{ll}
               A, & \mbox{if $k=T$} \\
               (k+1)\cdot A, & \mbox{if $k<T$} \\
              \end{array} \right.
\end{equation}

\noindent being $A$ the total number of actions performed by players.

In the worst case, $time(T,k,A) = (k+1)\cdot A$. $k$ can be considered a constant, therefore $time(T,k,A)$ is $O(A)$.

These limits show that the algorithm has a very small memory fingerprint and is not processing-intensive.
Our approach can be applied in real time and it can be implemented in any existing system.

\subsection{Training and Evaluation Capabilities} \label{sec:prtrain}

Given the low resource requirements of the system, several simultaneous parameterizations of the algorithm can be run online on the same game play data. Feedback such as penalties imposed to players by moderators or administrators allows performing automated learning by evaluating and adjusting the parameterizations and community quality functions in order to optimize the player ranking accuracy.

More interestingly, widely used optimization algorithms can be applied to parameter and function training by applying the system to extensive pre-recorded game play logs and evaluating its performance.

\subsection{Application to Minecraft} \label{sec:prmine}

Let $C$ be a Minecraft community, then the community quality value can be defined as:

\begin{equation}
Q(C) = \sum_{b \in C}b_{rarity}\cdot\frac{|b_{\mathrm{collected\_natural}}|}{|b_{\mathrm{collected\_placed\_other\_user}}|}
\end{equation}

This heuristic rewards players who collect more natural blocks than blocks placed by other players, and it penalizes players who collect blocks that were placed by other players and that, therefore, may be incurring of damage to artifacts. It should be noted that blocks removed by the same player that placed them do not alter the community quality.

Block rarity values are used as weights, as less common blocks that are placed by players have a higher chance of conforming a complex valuable player contraption.

Such a simple function is enough for our approaches to produce real-time Minecraft player rankings that highlight disruptive players who, at some moment, have started altering others contraptions.

It should be noted that this is a sample heuristic function and that it can obviously be greatly improved. One of the main advantages of our proposal is that a change in the heuristic function is enough to alter the criteria implicit in the player ranking.

\subsection{Application to a Discussion Forum} \label{sec:prdforum}

Let $C$ be a discussion forum community, then the community quality value can be defined as:

\begin{equation}
Q(C) = \frac{ \sum_{p \in C} \frac{p_{\mathrm{length}}}{  \frac{ p_{\mathrm{capital\_chars}} }{ p_{\mathrm{length}} } \cdot p_{\mathrm{forbidden\_words}} } }{|C|}
\end{equation}

The proposed quality function rewards long posts versus short posts. It also penalizes posts with high percentages of capital letters and forbidden words.

The definition of such a function is enough for our system to produce real-time forum users rankings. These rankings can be used to determine which users are disruptive and draw the moderator attention on them.

\subsection{Application to a Recommendation Site} \label{sec:prrsite}

Let $C$ be a recommendation site community, then the community quality value can be defined in function of the variances of each item ratings as:

\begin{equation}
Q(C) = \frac{ \sum_{i \in C} \sigma_i^2 } {|C|}
\end{equation}

Variances increases whenever outlying ratings are assigned to an item, therefore the proposed quality function reward new ratings that are close to the average rating for the item.

It should be noted that item ratings are subjective and distinct user sets may provide different consistent sets of ratings for items. In this case, users can be clustered using by considering the sets of user ratings they provided, so that users with the same likes pertain to the same sub-community.

The definition of such a simple function is enough for our technique to produce real-time recommendation site user rankings. These rankings can be used to determine user reputations, which can, in turn, alter the weights used for the calculation of item ratings.

\subsection{Application to the Collaborative Clustering Game} \label{sec:prccg}

Let $C$ be a collaborative clustering game. We defined a community quality function that reflects how close each color dot is to the same color dot centroid and how far it is from dots of different colors, as follows:

\begin{equation}
\begin{array}{l}
Q(C) = -\sum_{c \in C} \frac{\sum_{d \in c} dist(d,centroid(c))}{|c|} \\[0.3cm]
\hspace{0.3cm} +\sum_{c1,c2 \in C, c1 \neq c2} \frac{\sum_{d1 \in c1} \sum_{d2 \in c2} dist(d1,d2)}{|c1|\cdot|c2|} \\
              \end{array} 
\label{eq:ccg}
\end{equation}

\noindent being $centroid$ a function that returns the centroid of a color dots, and $dist$ a function that returns the euclidean distance between two positions.

Interestingly, the fact that the community quality function for this game is not heuristic but exact allows performing a quantitative comparison of the obtained results.

In the next section, we present, evaluate, and compare the rankings obtained from four parameterizations of our approach being applied to a game play of the collaborative clustering game.

\section{Approach Evaluation} \label{sec:comparison}

In order to evaluate the proposed approach for player ranking in collaborative games with no explicit objectives, we have run a 20-player play of the collaborative clustering game and we have collected and analyzed the game play data with the heuristic community quality function in Equation \ref{eq:ccg}.

Four classes of 5 players each have been asked to follow different behaviors during a sequence of 20 actions:
\begin{itemize}
\item The $F$ class are fair players.
\item The $f$ class were disruptive players who turned fair at some point.
\item The $d$ class were fair players who turned disruptive at some point.
\item Finally, the $D$ class are disruptive players.
\end{itemize}

While the $F$ and $D$ classes test the overall system performance and accuracy, the $f$ and $d$ classes test the system responsiveness capabilities.

We expected to obtain, at the end of the 20-action series, a ranking in which both the $F$ and the $f$ player classes were located in the upper half of the ranking and both the $d$ and the $D$ player classes were located in the lower half of the ranking.

Although it seems intuitive that an ideal ranking would list the $F$ player class in the interquartile $1$, the $f$ player class in the interquartile $2$ (as they are fair, but not as fair as the $F$ class players), the $d$ player class in the interquartile $3$ (as they are disruptive, but not as disruptive as the $D$ class players), and the $D$ player class in the interquartile $4$, it should be noted that the reason for the ranking is to determine whether players are being disruptive or fair at the current time, that is, at the end of the 20-action series. However, in the evaluation of our approaches we will give a slightly greater value to the fact that the player ranking makes a correct distinction between the $F$ and the $f$ player classes, or between the $D$ and the $d$ player classes.

\subsection{Approach Parameterizations} \label{sec:parameterizations}

We are going to evaluate the following four parameterizations of our contribution quality rating approach:

\begin{itemize}
\item $CQR_{\infty,0,\infty}$ returns the sum of all the deltas of a player's actions. This approach gives relevance to the behavior of players through time, and may be not enough responsive and accurate if the player's behavior changes.
\item $CQR_{8,0,8}$ returns the sum of the latest $8$ deltas of a player's actions. This approach should be more responsive than $CQR_{\infty,0,\infty}$, as in not more than $4$ actions in average it should be able to detect changes the behavior of players.
\item $CQR_{8,10,8}$ returns the sum of the latest $8$ deltas of a player's actions whose absolute value is greater than or equal to $10$. This approach might produce results that are intermediate between $CQR_{\infty,0,\infty}$ and $CQR_{8,0,8}$, as it takes into consideration more deltas, but, for that same reason, is less responsive.
\item Finally, $CQR_{8,10,4}$ returns the sum of the latest $8$ deltas of a player's actions whose absolute value is greater than or equal to $10$, however if the latest $4$ are all positive (resp. negative), all the negative (resp. positive) deltas are ignored. This approach should produce results that are better than $CQR_{8,0,8}$, as it takes into consideration more delta actions, but also better than $CQR_{8,10,8}$ and $CQR_{\infty,0,\infty}$, as it has a high responsiveness to behavior changes.
\end{itemize}

We now proceed to show the obtained results.

\begin{table*}[p]
\footnotesize
\begin{minipage}[b]{\linewidth}\centering
\setlength{\tabcolsep}{3.pt}
\centering
\caption{Comparison of four parameterizations of our player ranking approach: List of the action deltas and the contribution quality ratings for 20 players given 20 actions in the collaborative clustering game} \label{fig:comparison}
\vspace{2mm}
\begin{tabular}{c | r r r r r r r r r r r r r r r r r r r r | r r r r} \hline

id      & \multicolumn{20}{c}{action delta measures}                                                                            & \multicolumn{4}{c}{ranking algorithm} \\ \hline
        &     &     &     &     &     &     &     &     &     &     &     &     &     &     &     &     &     &     &     &     & \multicolumn{1}{c}{$CQR$}          & \multicolumn{1}{c}{$CQR$}          &  \multicolumn{1}{c}{$CQR$}          &  \multicolumn{1}{c}{$CQR$}          \\
        & 1   & 2   & 3   & 4   & 5   & 6   & 7   & 8   & 9   & 10  & 11  & 12  & 13  & 14  & 15  & 16  & 17  & 18  & 19  & 20  &     ${}_{\infty,0,\infty}$        &     ${}_{8,0,8}$                    &     ${}_{8,10,8}$                   &     ${}_{8,10,4}$                   \\ \hline
F1      & +05 & +10 & +10 & +20 & +05 & +40 & +07 & +02 & +10 & +12 & +05 & +15 & +01 & +42 & +18 & +26 & +20 & +35 & +08 & +09 & 300                              & 159                         & 178                          & 178                          \\
F2      & +12 & +16 & +36 & -41 & -56 & +15 & +06 & +08 & +01 & -19 & -37 & +33 & +36 & +42 & -15 & +51 & -32 & -34 & +12 & +04 & 38                               & 64                          & 93                           & 93                           \\
F3      & -01 & -04 & +06 & +12 & +19 & +05 & +10 & -04 & +03 & -10 & +40 & +30 & -14 & -12 & +32 & +37 & +11 & -14 & +02 & -15 & 133                              & 27                          & 55                           & 55                           \\
F4      & +10 & +40 & +25 & +05 & +35 & +45 & +45 & +05 & +01 & +01 & -02 & +10 & +15 & +05 & -01 & +10 & -05 & -15 & +30 & +30 & 289                              & 69                          & 170                          & 170                          \\
F5      & -03 & +12 & -05 & +06 & -06 & -15 & +21 & +07 & +06 & +09 & -17 & +13 & -16 & +05 & -32 & +20 & +13 & -05 & +22 & +24 & 59                               & 31                          & 27                           & 125                          \\ \hline
f1      & +75 & -12 & -15 & -32 & -46 & +05 & +14 & -22 & -57 & +24 & +12 & -03 & +25 & +12 & +04 & +01 & +12 & +14 & +01 & -03 & 9                                & 66                          & 20                           & 188                          \\
f2      & +15 & -18 & +12 & +14 & +17 & -12 & -43 & -24 & -22 & -37 & -36 & -32 & +14 & +25 & +24 & -12 & +24 & -03 & +24 & +24 & -46                              & 120                         & 91                           & 91                           \\
f3      & -12 & +24 & +28 & -14 & -16 & +25 & -14 & -32 & +12 & -22 & -28 & +14 & +12 & +17 & -09 & +03 & +05 & +12 & +02 & -02 & 5                                & 40                          & -15                          & 144                          \\
f4      & +14 & -42 & +17 & -11 & -15 & +02 & +04 & -18 & -21 & +20 & +14 & -12 & -14 & +23 & -26 & +29 & +12 & -14 & -01 & +32 & -7                               & 41                          & 30                           & 30                           \\
f5      & -12 & -41 & -45 & -22 & -14 & -17 & -19 & +04 & +06 & -12 & -16 & -14 & +02 & +18 & +16 & +19 & +23 & -03 & +05 & +12 & -110                             & 92                          & 46                           & 88                           \\ \hline
d1      & +14 & +17 & +37 & +27 & +54 & +41 & +12 & -02 & +16 & +17 & +12 & -14 & -17 & -24 & -52 & +11 & -32 & +02 & -14 & -06 & 99                               & -132                        & -130                         & -130                         \\
d2      & +32 & +34 & +12 & +05 & +09 & +14 & +27 & +14 & +25 & +15 & +14 & +25 & +12 & +06 & -14 & -25 & -05 & +07 & -14 & -16 & 177                              & -49                         & -3                           & -69                          \\
d3      & -27 & -29 & -15 & +25 & +28 & +12 & +16 & +27 & +45 & +32 & +29 & +31 & +12 & -14 & -17 & +16 & -03 & -08 & -17 & -19 & 124                              & -50                         & 21                           & 21                           \\
d4      & -42 & +22 & +24 & +17 & -29 & +12 & +05 & +09 & -07 & -29 & +14 & -34 & +03 & +08 & -12 & -05 & -15 & +12 & -16 & +21 & -42                              & -4                          & -59                          & -59                          \\
d5      & -14 & +26 & +17 & +05 & +26 & +26 & +39 & +12 & +17 & -34 & +16 & +25 & +02 & +09 & -16 & +25 & -32 & -12 & -15 & -24 & 98                               & -63                         & -33                          & -147                         \\ \hline
D1      & -27 & -12 & +15 & -04 & -16 & +02 & -18 & -31 & -12 & -14 & -19 & -05 & +12 & +05 & -12 & -15 & -10 & +09 & -24 & -01 & -177                             & -36                         & -94                          & -137                         \\
D2      & +04 & -14 & -16 & +15 & -13 & -23 & -17 & -07 & +04 & +05 & -17 & -29 & +12 & -04 & -14 & -23 & -11 & +10 & -12 & -16 & -166                             & -58                         & -83                          & -83                          \\
D3      & -17 & -14 & -12 & -16 & +05 & +12 & -16 & -21 & +03 & +13 & -22 & +25 & -06 & -16 & +26 & -12 & -37 & -15 & -18 & -03 & -141                             & -81                         & -69                          & -157                         \\
D4      & -05 & -25 & -21 & +05 & +16 & -12 & -15 & -13 & +24 & -17 & -14 & +09 & -24 & -12 & -16 & +14 & -13 & +02 & -21 & -34 & -172                             & -104                        & -120                         & -120                         \\
D5      & -36 & +45 & -23 & -25 & -14 & -51 & +21 & +12 & -14 & -19 & -24 & -16 & -04 & -05 & +02 & +03 & -15 & +12 & -14 & -42 & -207                             & -63                         & -132                         & -132                         \\ \hline
\end{tabular}
\end{minipage}

\vspace{3mm}

\centering
\caption{Rankings obtained from the application of the four parameterizations of our player ranking approach} \label{fig:rankings}
\begin{minipage}[b]{0.24\linewidth}\centering
\vspace{2mm}
\hspace{0.4cm}
\setlength{\tabcolsep}{3.pt}
\begin{tabular}{c | c | r} \hline
\multicolumn{3}{c}{$CQR_{\infty,0,\infty}$} \\ \hline
rank & id & value \\ \hline
1  & F1 & 300 \\
2  & F4 & 289 \\
3  & d2 & 177 \\
4  & F3 & 133 \\
5  & d3 & 124 \\ \hline
6  & d1 & 99 \\
7  & d5 & 98 \\
8  & F5 & 59 \\
9  & F2 & 38 \\
10 & f1 & 9 \\ \hline
11 & f3 & 5 \\
12 & f4 & -7 \\
13 & f2 & -46 \\
14 & d4 & -42 \\
15 & f5 & -110 \\ \hline
16 & D3 & -141 \\
17 & D2 & -166 \\
18 & D4 & -172 \\
19 & D1 & -177 \\
20 & D5 & -207 \\ \hline
\end{tabular}
\end{minipage}
\begin{minipage}[b]{0.24\linewidth}
\centering
\setlength{\tabcolsep}{3.pt}
\begin{tabular}{c | c | r} \hline
\multicolumn{3}{c}{$CQR_{8,0,8}$} \\ \hline
rank & id & value \\ \hline
1  & F1 & 159 \\
2  & f2 & 120 \\
3  & f5 & 92  \\
4  & F4 & 69  \\
5  & f1 & 66  \\ \hline
6  & F2 & 64 \\
7  & f4 & 41 \\
8  & f3 & 40 \\
9  & F5 & 31 \\
10 & F3 & 27\\ \hline
11 & d4 & -4\\
12 & D1 & -36\\
13 & d2 & -49 \\
14 & d3 & -50 \\
15 & D2 & -58  \\ \hline
16 & D5 & -63  \\
17 & d5 & -63  \\
18 & D3 & -81  \\
19 & D4 & -104 \\
20 & d1 & -132 \\ \hline
\end{tabular}
\end{minipage}
\begin{minipage}[b]{0.24\linewidth}\centering
\setlength{\tabcolsep}{3.pt}
\begin{tabular}{c | c | r} \hline
\multicolumn{3}{c}{$CQR_{8,10,8}$} \\ \hline
rank & id & value \\ \hline
1  & F1 & 178 \\
2  & F4 & 170 \\
3  & F2 & 93  \\
4  & f2 & 91  \\
5  & F3 & 55  \\ \hline
6  & f5 & 46 \\
7  & f4 & 30 \\
8  & F5 & 27 \\
9  & d3 & 21 \\
10 & f1 & 20\\ \hline
11 & d2 & -3\\
12 & f3 & -15\\
13 & d5 & -33 \\
14 & d4 & -59 \\
15 & D3 & -69  \\ \hline
16 & D2 & -83  \\
17 & D1 & -94  \\
18 & D4 & -120 \\
19 & d1 & -130 \\
20 & D5 & -132 \\ \hline
\end{tabular}
\end{minipage}
\begin{minipage}[b]{0.24\linewidth}
\centering
\setlength{\tabcolsep}{3.pt}
\begin{tabular}{c | c | r} \hline
\multicolumn{3}{c}{$CQR_{8,10,4}$} \\ \hline
rank & id & value \\ \hline
1  & f1 & 188 \\
2  & F1 & 178 \\
3  & F4 & 170 \\
4  & f3 & 144 \\
5  & F5 & 125 \\ \hline
6  & F2 & 93 \\
7  & f2 & 91 \\
8  & f5 & 88 \\
9  & F3 & 55 \\
10 & f4 & 30\\ \hline
11 & d3 & 21\\
12 & d4 & -59\\
13 & d2 & -69 \\
14 & D2 & -83 \\
15 & D4 & -120 \\ \hline
16 & d1 & -130 \\
17 & D5 & -132 \\
18 & D1 & -137 \\
19 & d5 & -147 \\
20 & D3 & -157 \\ \hline
\end{tabular}
\end{minipage}
\end{table*}

\begin{table}[h]
\begin{minipage}[b]{\linewidth}
\setlength{\tabcolsep}{3.pt}
\centering
\caption{Player class to interquartile correspondence values for ranking evaluation} \label{fig:values}
\begin{tabular}{c | r r r r} \hline
      & \multicolumn{4}{c}{Interquartile}   \\ \hline
class & I   & II  & III & IV   \\ \hline
F     & 6   & 4   & -10 & -25   \\
f     & 4   & 6   & -4  & -10   \\
d     & -10 & -4  & 6   & 4   \\
D     & -25 & -10 & 4   & 6   \\ \hline
\end{tabular}
\setlength{\tabcolsep}{3.pt}
\centering
\caption{Quantitative comparison of the four parameterizations of the approach using the results in Table \ref{fig:rankings} and the correspondence values in Table \ref{fig:values}} \label{fig:compsummary}
\begin{tabular}{l | c} \hline
parameterization           & value \\ \hline
$CQR_{\infty,0,\infty}$  & 34  \\
$CQR_{8,0,8}$            & 100  \\
$CQR_{8,10,8}$           & 92    \\
$CQR_{8,10,4}$           & 104    \\ \hline
\end{tabular}
\end{minipage}
\end{table}

\subsection{Results and Discussion} \label{sec:obtainedresults}

Table \ref{fig:comparison} shows the delta values associated to each of the 20 actions of the 20 players. Positive values reflect fair actions, negative values reflect negative actions. Greater absolute values reflect actions with a greater impact in the community.
Table \ref{fig:rankings} shows the four rankings corresponding to the four parameterizations.

From the analysis of Table \ref{fig:rankings}, it can be observed that our predictions on the results were accurate:
\begin{itemize}
\item $CQR_{\infty,0,\infty}$ got many disruptive and fair players intermixed due to the fact that it considers all the deltas of player actions and therefore is not responsive toward changes in the behavior of players.
\item $CQR_{8,0,8}$ is highly responsive and has managed to classify in the upper half of the ranking all the players who ended up acting fairly, and in the lower half of the ranking all the players who ended up acting disruptively. Some intermixing of the $F$ and the $f$ player classes and of the $D$ and the $d$ player classes can be observed, as this technique is not able to consider past action deltas.
\item $CQR_{8,10,8}$, in contrast, takes into account older action deltas by ignoring some of the most recent low value deltas. However, it loses responsiveness as a side effect of this, which causes a slight intermixing of the players classes in the ranking.
\item Finally, $CQR_{8,10,4}$ produces a correct ranking. This is due to the fact that the extra past action deltas that are considered is balanced with the extra responsiveness filter.
\end{itemize}

In order to quantitatively compare the results procuded by the different parameterization of our approach, we have assigned values to the correspondence between the ranking interquartiles and the player classes in the ranking. Table \ref{fig:values} shows these correspondence values.

Table \ref{fig:compsummary} presents the quantitative comparison on the correspondence between ranking interquartiles and player classes for the four approaches using the correspondence values in Table \ref{fig:values}. As it can be observed in this table, $CQR_{\infty,0,\infty}$ has the lowest correspondence, $CQR_{8,0,8}$ greatly improves it, $CQR_{8,10,8}$ offers a slight decrease in accuracy, and $CQR_{8,10,4}$ improves it back again and over $CQR_{8,0,8}$. This, again, reflects our predictions.

Overall, the rankings performed by the $CQR_{8,0,8}$ and $CQR_{8,10,4}$ parameterizations of our approach are very good approximations of the expected ranking and similar parameterizations can be used to help in the management of collaborative communities and games.

In the next section, we present our conclusions and the future work that derives from our research.

\section{Conclusions and Future Work} \label{sec:concfw}

Player ranking in collaborative games allows determining whether a player is contributing to a community or harming it. Player rankings are used in games or communities to reward contributing players with perks or to punish disruptive players. These perks or penalizations usually modulate the impact of the player's actions in the community.

When the game has clear shared objectives, it is easy to determine how much a player is contributing to the community, just by measuring how much the player's actions contribute to the objectives completion.

We have proposed a parameterizable approach for real-time player ranking in collaborative games with no explicit objectives that produce player rankings from the application of heuristic community quality functions.

Our approach has been successfully tested with a game play of the collaborative clustering game. In this experiment, the players were ranked correctly according to their intended behavior, which demonstrates that our approach provides good approximations to player rankings.

In the future, we plan to research on automated learning of community quality functions and contribution quality rating expression parameters by analyzing complex tagged game play logs, in order to automatically produce ranking mechanisms for collaborative games and communities.

\bibliographystyle{plain}
\bibliography{doc}

\end{document}